\documentclass[twocolumn,showpacs,preprintnumbers,amsmath,amssymb]{revtex4}

\usepackage{color,graphicx}
\usepackage{dcolumn}
\usepackage{bm}

\begin{document}
[{\small A.\ E. Motter and Z. Toroczkai, Chaos {\bf 17}, 026101 (2007)}]

\title{Optimization in Networks}

\author{Adilson E. Motter}
\email[E-mail address:]{motter@northwestern.edu}
\affiliation{Department of Physics and Astronomy and Northwestern Institute on Complex Systems, Northwestern University, Evanston, IL 60208, USA} \author{Zoltan Toroczkai}
\email[E-mail address:]{toro@nd.edu}
\affiliation{Department of Physics, University of Notre Dame, Notre Dame, IN 46556, USA}

\date{\today}

\begin{abstract}
The recent surge in the network modeling of complex systems has set the stage for a 
new era in the study of fundamental and applied aspects of optimization in collective 
behavior. This Focus Issue presents an  extended view of the state of the art in this 
field and includes articles from a large variety of domains where optimization manifests itself, 
including physical, biological, social, and technological networked systems.
\end{abstract} 

\maketitle

{\bf \noindent One of the broadest areas of research, optimization has a very 
long history. It comprises the variational principles in physics and engineering, 
the survival-of-the-fittest principles that pervade biology and economics, the founding 
hypotheses of numerous computer algorithms, and the frameworks for addressing the 
improvement of efficiency in various contexts. Whether a fact or a goal, a natural 
process or a man-made system, the apparently ubiquitous striving for optimization 
generates continuing appeal among researchers. But what is new about optimization in 
{\it networked} systems?}

\smallskip

\begin{figure}[b]
\includegraphics[width=8.5cm]{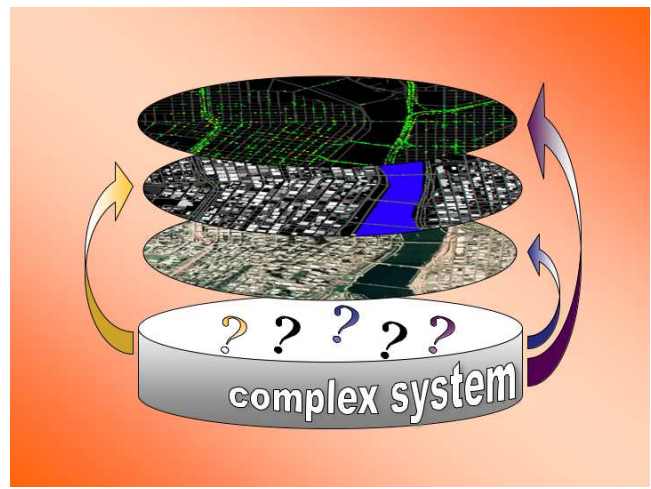}
\caption{Illustration of levels of abstraction represented as networks
that underlie a complex system, in this case an urban area.
What network is relevant to an optimization problem depends on the specific
research question at hand. The inlay images are courtesy of the TRANSIMS and EpiSims teams at LANL and
Virginia Bioinformatics Institute ({\tt http://ndssl.vbi.vt.edu}) \cite{Eubank,barret}. 
\label{fig:fig1}} \end{figure}

Real-world systems do not operate isolated from each other. While a neuron can be studied 
in a laboratory setting, it has not evolved to work independently of the 
activity of other neurons nor has the brain evolved to work independently from the organism. 
  In a hierarchy of scales, many systems are formed by the 
interconnection of subsystems that may have different (or even opposing) optimization goals 
than the global system which they are part of. Expectedly, 
the structure of these interconnections
will influence the global performance and hence complex network research
\cite{WS98, science}
 is a key ingredient
for studying optimal system behavior  (see Fig. \ref{fig:fig1}).  Not surprisingly, various structural and dynamical 
network properties have been explicitly related to the optimization of specific functions 
(see, e.g., Refs. \cite{Ferrer,Nishikawa,Guimera} for early works and Refs. 
\cite{rev1,rev2,rev3,rev4,rev5} for recent reviews).  In this context, there are entire 
classes of problems, ranging from epidemic spreading \cite{Eubank} to the control of 
cascading failures \cite{Motter}, which are naturally defined as extremization problems. 
Others, such as the unexpected robustness observed in some systems, involve no {\it a priory} 
optimization conditions and yet reveal enhanced properties shaped by the evolution of the 
system.

Optimal behavior is most often connected to a function that the system performs.
In numerous cases the function is multi-variate or multi-faceted. For example, the power grid
has as its main role the transport of electric energy while minimizing generation and distribution
costs and maximizing at the same time reliability and quality of service. `How network structure
influences the global performance of such systems' is probably the question that is posed most frequently in network research. Conversely, the notion of network acts as a unifying theme 
in systems optimization. Indeed, with the 
increasing abundance of empirical and theoretical results, numerous optimization questions 
involving apparently unrelated systems can be addressed using a common formalism of complex 
networks of interacting units.  

Traditionally, optimization has a strict mathematical definition, which refers to obtaining {\em the} solutions
that strictly extremize a well-defined functional. Here we adopt a looser definition of the word by extending it 
to include a {\em tendency} of the system to improve its behavior as a result of a selection pressure naturally
or artificially imposed. Many real-world problems are complex, with a very large parameter
space. Accordingly, most quests for finding ``the'' optimizing or ``the'' best solution is doomed to failure or 
not to be realistic. 
Finding a ``good enough solution'' or a ``better solution'', perhaps even in an iterative manner,
becomes, however, a workable alternative.  Processes in nature follow this path. 
Nature has evolved biological systems  under pressure towards increasingly optimal behavior and much can be learned by studying the behavior of these systems. 
Despite the complexity of the problems, the original work on scale-free networks already suggested that there could be  some general principles in network optimization \cite{science}: many realistic networks, including the 
Internet and collaboration networks, tend to grow and evolve much in the same way as some 
biological systems do. If selection is important in biology, then it is expected to be 
important in other evolving systems too. \\

\noindent {\bf Broad View of Network Optimization}\\

There are fundamentally four major types of optimization problems related to networked systems
(the constraints are considered to be implicit in the functional): \\

\noindent Type I - {\em Structural Optimization}. Find a graph $G(V,E)$ where $V$ is the set of nodes and $E$
is the set of edges  which extremize  a given structural functional 
${\bf F}[G]$.\\

\noindent Type II - {\em Dynamics Optimization on Static Graphs}. For a given graph $G(V,E)$ and a {\em
dynamical system} $\mathbf{\Phi}$ on $G$,
\begin{equation}
\mathbf{\Phi}({\bf x},{\bf \dot{x}},\ldots,\{\alpha\},t) = \mathbf{0} \label{dyn},
\end{equation}
find the values of the parameters $\{\alpha\}$ which extremize a global functional $\mathbf{F}[\mathbf{\Phi}]$
of  the dynamics $\mathbf{\Phi}$. The variables $\mathbf{x}$ are quantities associated with 
properties of the nodes and edges in the network.\\

\noindent Type III - {\em Structural Optimization for Dynamics}. Given the dynamical system (\ref{dyn}) and a
set of parameters $\{\alpha\}$, find a graph $G(V,E)$ for which a global functional $\mathbf{F}[\mathbf{\Phi}]$
of  the dynamics $\mathbf{\Phi}$ is extremized.\\

\noindent Type IV - {\em Dynamics-driven Network Optimization}.  If the graph of the network evolves in time
(i.e., $G(V,E) = G(V,E,t)$), either through an independent dynamics or through coupling to 
the dynamics in (\ref{dyn}), find the values of the parameters 
$\{\alpha\}$ for which a global functional $\mathbf{F}[G,\mathbf{\Phi}]$ of the dynamics $\mathbf{\Phi}$ {\em and} of
the graph $G(V,E,t)$ is extremized.\\

Type I is a purely graph theoretic problem in that one looks for structures that have some specified properties.
For example, given a fixed degree sequence on $N$ nodes, construct a graph that minimizes the diameter.  Problems
involving optimal assignment of edge-weights and -directionality also belong to this class.  Type II is a 
``flow extremization'' problem.  For example, given a roadway network, what should be the speed limit for cars on every street such that jamming is minimized? Type III commonly occurs in design problems: given a
flow dynamics, such as packet flow in packet switched networks, find the graph structures optimal for
information throughput.  Other important examples include the optimization of synchronous and coherent behavior.
Type IV is also common, though sometimes very difficult to solve because properties of
both graph structure and dynamics are allowed to change. This is also the most relevant case to the study of
emergent properties in evolving systems.
Robustness and vulnerability problems fall into this class when the flow through the network can change the
structure, which in turn changes the flow. Prime examples of this are cascading failures in networked systems,
such as power grids (see the article by Dobson {\em et al.} in this issue). 

Optimization in complex networks is thus of broad significance, 
incorporating static and dynamical properties and serving as an instrument to analyze and control 
the evolution and function of both natural and engineered systems.\\

\noindent {\bf This Focus Issue}\\

This Focus Issue brings together contributions on network structure and dynamics, 
with emphasis on optimization problems and their applications to infrastructure and 
biological systems. Key topics discussed include the optimization in the evolution 
and functioning of biological systems, optimization and cost balance analysis in the 
design of infrastructure networks, and optimization principles emerging from the interplay
between network structure and dynamics.

In the context of technological and infrastructure networks, Danila {\it et al.} 
\cite{Danila_etal} consider routing optimization in network transport,  Dobson 
{\it et al.} \cite{Dobson_etal} discuss how competition between efficiency and robustness 
leads to a SOC-based model for the power-grid dynamics,  Guclu {\it et al.} \cite{Guclu_etal} 
study how fluctuations and synchrony in distributed processing networks relate to the 
network structure,  Gulbahce \cite{Gulbahce} addresses the optimization of jamming on 
gradient networks, while Teuscher \cite{Teuscher} analyzes the impact of performance 
metrics in network-on-chip designs.

In the context of biological networks, Almaas \cite{Almaas} studies metabolic flux 
patterns derived from flux-balance optimization assumptions, Balcan and Erzan 
\cite{Balcan_Erzan} discuss a statistical physics description of content-based networks 
which can serve as models for gene regulatory networks,  Mahmoudi {\it et al.} 
\cite{Mahmoudi_etal} consider the propagation of external regulatory information and 
asynchronous dynamics in random Boolean networks, and Riecke {\it et al.} \cite{Riecke_etal} 
study a rich variety of dynamical states in the activity of small-world networks of 
excitable neurons.

Other dynamical processes are also considered. Barrat {\it et al.} \cite{Barrat_etal} 
address the emergence of consensus in linguistic conventions,  Bogacz {\it et al.} 
\cite{Bogacz_etal} consider condensation and far from equilibrium dynamics on networks,  
and Freire {\it et al.} \cite{Freire_etal} analyze synchronization and complex 
spatio-temporal patterns in networks of cellular automata.

More related to the structural properties of the networks,  Bianconi \cite{Bianconi} 
studies how the degree distribution follows from the extremization of a free-energy 
function, Kim and Kahng \cite{Kim_Kahng} derive spectral densities for an important 
class of weighted complex networks, Kim {\it et al.} \cite{Kim_etal} analyze the fractal 
properties of complex networks, and Minnhagen and Bernhardsson \cite {Minnhagen_Bernhardsson} 
study how the degree distribution relates to maximization of information.

Summing up, optimization of performance and robustness is a common property of naturally 
evolved systems and is a desirable property in most man-made systems. There is now 
increasing evidence that the functioning of complex systems as diverse as cellular metabolism 
and power grids lies deep in the properties of underlying complex networks. This 
evidence has generated increasing interest on dynamical processes in complex networks 
and on how the interplay between these processes and network structure influences 
the performance and robustness of the system. Notably, established areas such as resource 
management, epidemic spreading, communication processes, synchronization dynamics, 
cellular biology, and cascading failures are at the leading edge of the current research 
on network optimization. We hope that this Focus Issue will provide the reader with an up 
to date overview of this exciting area of research.\\

{\bf \noindent Acknowledgments.}
We would like to thank the Center for Nonlinear Studies and the Complex Systems 
Group at Los Alamos National Laboratory, particularly Robert Ecke, Adam Shipman, 
Donald Thompson, and Elissa Vigil, for sponsoring and coordinating the Workshop 
that led to this Focus Issue.  We specially thank Chaos Editorial Office members 
Janis Bennett and David Campbell for their prompt assistance and all the authors 
for their invaluable contributions.

\end{document}